**The Critical Brain Hypothesis? Meet The Metastable Brain~Mind**

J.A. Scott Kelso
Center for Complex Systems & Brain Sciences,
Florida Atlantic University,
Boca Raton, Florida, USA
&
Intelligent Systems Research Centre,
Ulster University—Magee Campus,
Derry~Londonderry, N. Ireland

In contrast to, or perhaps along with, the criticality hypothesis (https://www.quantamagazine.org/a-physical-theory-for-when-the-brain-performs-best-20230131/) of John Beggs and Dietmar Plenz (both of whose work I greatly respect, e.g., Kelso, 2014) in which the brain "optimizes its information processing" in between the ordered, strongly coupled and disordered, weakly coupled régimes of neural function-- by self-tuning to a critical point, a more complex picture exists which says:

1) the functioning brain has a vast repertoire of coexisting tendencies/dispositions for regions of the brain to integrate and segregate at the same time. This assures the brain a vital mix of both flexibility and stability;

2) The degree to which one or the other tendency may dominate (or one might say 'the balance' between these tendencies) will depend on a number of factors, including age, dementia, task, memory, functional connectivity, propagation delays, circulating neurohormones, etc., etc. (see, e.g., Alderson, et al., 2020)

3) The picture below shows that a near infinite, uncountable number of *metastable* tendencies exist between order (stable) and disorder (uncoupled). This happens when all the states have disappeared (see **A**). The outcome is that rather than teetering between order and randomness, seizure and inactivity, *the brain (or more properly, the brain~environment system) lives in a sea of metastability.* The secret, as the poet Robert Frost said, sits in the middle and knows. This middle is more of a niche for the brain to inhabit than a 'sweet spot' or a boundary between order and randomness favored by the critical brain hypothesis.

Consider the following figure.

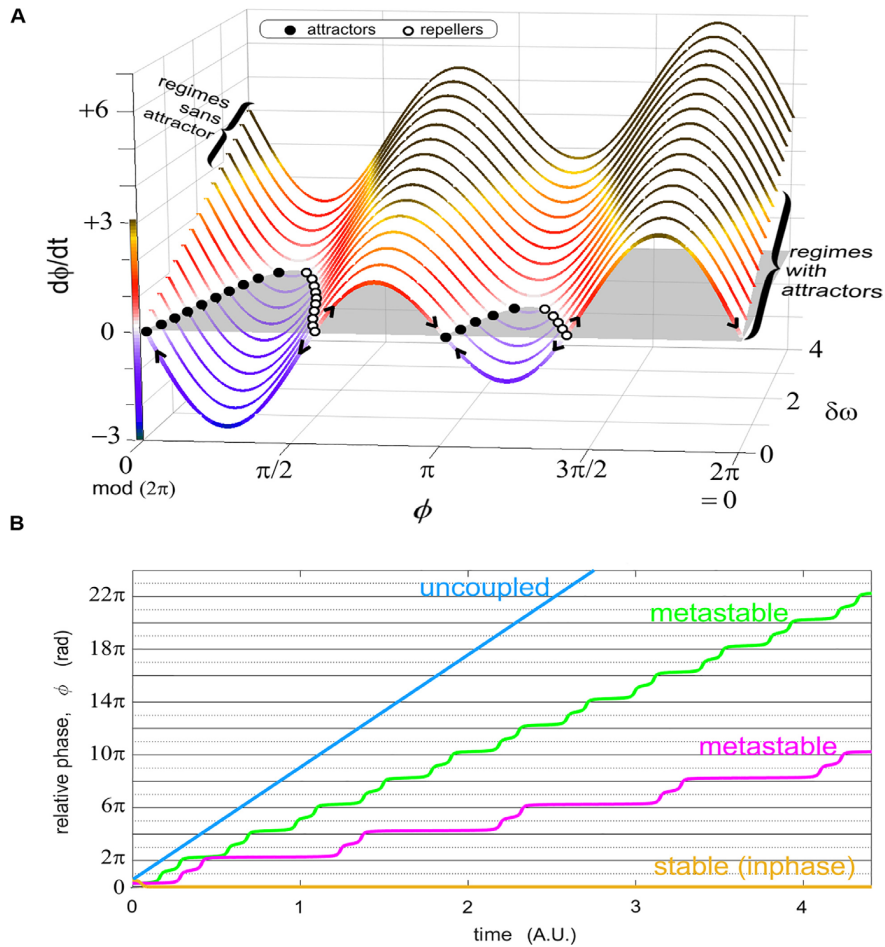

Figure 1. **A.** Shown is the phase portrait of the so-called 'extended HKB model' of Coordination Dynamics which has two control parameters, the *coupling strength* between the components, and a *heterogeneity parameter* representing *differences* between the components. Looking from bottom to top, the system has two stable equilibria or fixed points (filled circles) representing *states of coordination*, one near in-phase and one near antiphase. In other words, the system is *bi-,* or in general *multi-stable*, and corresponds to *régimes with attractors*. Notice as parameters are changed (e.g., weaker coupling, more heterogeneity) stable and unstable (repelling) fixed points (open circles) collide in a *saddle node bifurcation*. First the state near antiphase shifts and disappears, leaving only a stable fixed point near in-phase. Predicted features of *nonequilibrium phase transitions* have been observed in both large scale brain recordings (MEG, EEG) and behavior, as the system approaches criticality. Non-equilibrium phase transitions (NEPT) are not the same as *self-organized criticality* where the system tunes itself to critical or tipping points. Rather, in NEPT the system's behavior is subject to *control parameters* such as the concentration of neuromodulators that lead it through critical points (cf. Haken, 1977/1983). Notice that after all the fixed points, stable and unstable, have disappeared (*régimes sans attracteurs*) there are still tendencies to where the fixed points once were (sometimes referred to as 'ghost states' in brain science). In Coordination Dynamics, this is called *metastability* (meta meaning beyond; Bressler & Kelso, 2016; Kelso, 2012; Tognoli & Kelso, 2014). That is, tendencies for the components to remain coordinated coexist at the same time as tendencies for the components to function independently. Strictly speaking, in the metastable, post-critical régime there are no longer any states, stable or unstable, coordinated or uncoordinated, only tendencies to where the states once were. **B.** The corresponding time evolution of the relative phase order parameter/collective variable shows that a vast region of metastable coordination exists 'in between' stable, ordered coordination states (here the flat yellow line shown only for in-phase coordination) and the uncoupled situation where the parts are not ordered and independent. In that case, the order parameter relative phase simply grows continually in time (blue). Notice the transient *dwell* near formerly stable coordination states and *escape* from them, the dwell time duration depending on the proximity to stable coordination states. One can readily see, for example, that the system hangs around longer and longer the closer it is to (here stable) in-phase coordination (magenta). The system may appear to be in a stable state when in fact it is in a long transient. If this picture is correct, and there is much evidence to think that it is, rather than (or at least in addition to) finely tuning itself to a critical point between chaos and order, the brain lives in a virtual *sea of metastability*. It's not so much a border or a boundary between order and disorder, but a niche in between them that the brain inhabits.

Several other points are worthy of mention. Though simple in appearance, the core of the extended HKB model consists of nonlinearly coupled nonlinear oscillators. An argument can and has been made by biophysicists such as Morowitz, Iberall, Yates and others that the nonlinear oscillator is a basic unit of life. Fundamentally the extended HKB equation is a basic equation that describes how different component parts come together as a collective--regardless of whether it is coordination between the components of an organism, organisms themselves and organisms and their environment (Kelso, 2022). Whereas the critical brain hypothesis claims that by operating near critical points the cortex can perform "optimal information processing", the metastable brain~mind[1] may be seen as *creating* information. Starting at the top of Figure A, for example, where no states exist, it is easy to see that as a result of parameter changes, coordination states (filled circles) appear spontaneously. It does not require a huge step of imagination to conceive of the emerging in-phase and anti-phase states as binary state equivalents of 1 and 0. Notice also, that in the metastable régime, close to fixed points, noise can play a constructive role, 'kicking' the system into stable states of coordination.

On a biological level, such functional coordination states and tendencies are relational and meaningful. They are observed as patterns of relative phase, ubiquitous at multiple levels of the nervous system—from rhythms in the cortex of humans to "central pattern generators" throughout the animal kingdom (Grillner, 2018). What if, instead of zeroes and ones, computations were based on collective coordination variables having the characteristic of a phase defined in the space of zero to $2\pi$? Such a metastable computer would be based on known properties of coordination or *collective behavior of neurons in the brain*, not simply the on-off threshold-like elements/"neurons" of current computational neuroscience. This would usher in a biological Dynamical Neuroscience[2] not only with far greater potential computational capabilities but also with far greater biological realism (see, e.g., Izhikevich, 2007). According to Coordination Dynamics, that realism complements the 'sweet spot' theory of brain criticality with a brain~mind that functions naturally in an immense sea of metastability.

**Acknowledgments**

I am grateful for the comments and insights of Drs. Guillaume Dumas, David Engstrøm, Fran Hancock, Viktor Jirsa, Brian Josephson, Joanna Rączaszek-Leonardi, Aliza Sloan and Mengsen Zhang. Any errors herein, however, are mine.

**Footnotes**

[1] The squiggle or tilde symbol reflects the complementary, non-dualist nature of brain and mind whose scientific grounding is metastable coordination dynamics (Kelso & Engstrøm, 2006).

[2] In fact, the term "Dynamical Neuroscience" refers to a series of 20 conferences (1993-2012) organized by D. Glanzman and the author sponsored by the National Institute of Mental Health. The list is available from the author.